\documentstyle[epsf]{mn}
\begin{document}

\title[Very high energy gamma radiation of the unshocked wind of the Crab pulsar]
{Very high energy gamma radiation 
associated with the unshocked  wind
of the Crab pulsar}
\author[S.V. Bogovalov \&  F.A. Aharonian]
{S.V. Bogovalov,$^1$ F.A. Aharonian$^2$\thanks{E-mail: Felix.Aharonian@mpi-hd.mpg.de}\\
$^1$ Astrophysics Institute at the Moscow 
Engineering Physics Institute, Kashirskoe Shosse 31, Moscow 115409, Russia\\ 
$^2$Max Planck Institut f\"ur Kernphysik,
Postfach 103980, D-69029 Heidelberg, Germany}
\date{Accepted 1999 November 10. Received  1999 November 10; 
in original form  1999 March 4}

\maketitle
\begin{abstract}
We show that the relativistic wind of the Crab pulsar, which is commonly
thought  to be invisible in the region upstream of the termination shock
at $r \leq r_{\rm S} \sim 0.1 \, \rm pc$, in fact could be directly
observed through its inverse Compton  (IC) $\gamma$-ray emission. This
radiation is caused by illumination of the wind by low-frequency photons
emitted by the pulsar, and  consists of two,
{\it pulsed} and {\it  unpulsed},  components associated with the nonthermal
(pulsed) and thermal (unpulsed) low energy radiation 
of the pulsar, respectively.
These  two components of $\gamma$-radiation  have distinct
spectral characteristics,  which depend essentially
on the site  of formation of the kinetic-energy-dominated wind, as well as
on the  Lorentz factor and the geometry of propagation of the wind.
Thus, the search for such specific  radiation components
in the spectrum of the Crab Nebula
can provide unique information about the unshocked pulsar wind
that  is not accessible at  other wavelengths.
In particular, we show that the comparison of the calculated flux  of
the unpulsed  IC emission with the measured $\gamma$-ray flux of
the Crab Nebula excludes  the possibility of formation  of a 
kinetic-energy-dominated wind within
5 light cylinder radii of the pulsar, $R_{\rm w} \geq 5 R_{\rm L}$.
The analysis of  the pulsed IC emission,
calculated under reasonable assumptions concerning the production site and
angular distribution of the optical pulsed radiation,
yields even tighter restrictions, namely $R_{\rm w} \geq 30 R_ {\rm L}$.
\end{abstract}

\section{Introduction}
\label{intro}

The Crab pulsar is a powerful nonthermal machine,  accelerating
plasma  in the form of a relativistic wind that carries
off most of the  rotational energy of the pulsar.

At a distance  of about $r=r_{\rm S} \sim 0.1 \, \rm pc$
the wind  is terminated by a standing reverse shock,  which
accelerates the electrons up to energies $10^{15} \, \rm eV$,
and randomizes their pitch angles \cite{rees,kennel}.
This results in  formation of a bright synchrotron  source in the
region downstream of the shock.
The synchrotron  radiation of the Crab Nebula
is well studied in a very broad frequency range, from
radio to hard X-rays. Its general spectral
and spatial characteristics are satisfactorily explained
by the relativistic magnetohydrodynamics (MHD) 
model of Kennel \& Coroniti \shortcite{kennel}.
Remarkably, the latter  provides also a reasonable explanation,
even in its simplified (spherically symmetric)  form, 
for the detected very high energy (VHE) $\gamma$-rays
as a result of inverse Compton (IC)
scattering of relativistic electrons in the ambient
low-frequency photon  fields. This implies that the study of the
TeV IC radiation of the Crab Nebula,  combined with synchrotron
X-ray emission, can yield  unambiguous information about the relativistic
electrons and the nebular magnetic field  in the downstream region
of the shock  \cite{jager,stepanian,atoyan,hillas}.

Although very important, this information unfortunately does not tell
much about the origin and characteristics of the wind, i.e. about the
region  between the pulsar magnetosphere  and the shock.
It is generally believed that this region, where almost the whole
rotational energy of the pulsar is somehow released in the
form of kinetic energy of  the wind, cannot be directly observed.
This has a simple explanation. Although the wind electrons may have
an energy as large as $\sim 10^{13}$ eV,  they move  together with the
magnetic field  and thus do not emit synchrotron radiation.
This explains the fact that 
 the region upstream of the shock is {\it underluminous} 
\cite{kennel}. However, this statement is valid
only for the synchrotron radiation of the wind. In fact,
the wind could be {\it directly}  observed through its IC radiation.
Indeed, the IC $\gamma$-radiation   of  the wind electrons  is
unavoidable because of  the illumination of the wind by external low-energy
photons of different origin.
There are three isotropic photon field populations that
contribute effectively  to the  production of IC
$\gamma$-radiation  of the nebula in the downstream region:
the nonthermal (synchrotron) and thermal (dust)  radiation of the Crab
Nebula itself,
and the 2.7 K microwave background radiation (Atoyan \& Aharonian 1996).
Formally, all these photon fields could also serve as a target for
the IC scattering of the wind before the shock.
However, the fluxes of  $\gamma$-rays  emitted through these channels
appear to be well below the sensitivity threshold of $\gamma$-ray
telescopes. Meanwhile, in the pulsar vicinity, namely within
approximately  100 light cylinders, the efficiency of production of
IC $\gamma$-rays dramatically increases because of  the existence of intense
low energy radiation of the pulsar itself.
The radiation  consists of two, pulsed and unpulsed components,
the latter being modulated at the period of the pulsar.

In this paper we show that at certain circumstances  concerning the
position  of formation of the particle dominated wind, 
the geometry of the flow and
the Lorentz-factor of the bulk motion  of the wind, the fluxes of
the IC $\gamma$-radiation of the wind could be sufficiently high
to enable  detection by present and forthcoming
space-borne and ground-based $\gamma$-ray telescopes.
Moreover, the distinct spectral features of this radiation could
allow  effective  separation of  the ``wind'' component of radiation
from the  heavy background,   which consists of
unpulsed radiation of the Crab Nebula at very high (TeV) energies
and pulsed radiation at low (GeV) energies.
We argue, that even upper limits obtained in such a study could provide
unique information about the origin of the pulsar wind.

\section{Characteristics of the wind}

\subsection{The total particle ejection rate of the wind}

The wind from the Crab pulsar carries away most of the 
energy of rotation of the pulsar. The energy released in the form of 
electromagnetic emission of the pulsar, 
which peaks at  gamma-ray energies, 
does not exceed 1 per cent of the total rotational losses \cite{arons95},
therefore it can be neglected in the energy balance of the wind.

The fluxes of the energy and the angular momentum of the wind consist 
of two parts. One of them corresponds to the matter and another 
corresponds to the electromagnetic field.
The total flux of the kinetic energy of particles 
can be presented as
\begin{equation}
\dot E_{\rm kin}=\dot N mc^{2} <\gamma_{w}>,
\label{eq1}
\end{equation}
while the flux of the angular momentum of the matter 
is equal to
\begin{equation}
\dot L_{\rm kin} =\dot N m <r\gamma_{w} v_{\varphi}>.
\label{eq2}
\end{equation}
Here $\dot N$ is the total rate of particle ejection from the pulsar
magnetosphere, $<\gamma_{\rm w}>$ is the average Lorentz-factor of the 
wind,  $r$ is the distance to the axis of rotation  $v_{\varphi}$
is  the component of the velocity of plasma 
propagation in the direction of rotation.
Hereafter we  assume that the wind consists of only 
electrons and positrons.
The total rate of particle ejection can be estimated 
within the models $e^{\pm}$ pair production in the pulsar 
magnetosphere:
\begin{equation}
\dot N_{\rm gap}= n_{\pm}cS_{\rm cap} \lambda,
\end{equation}
where $S_{cap}=2\pi R_{*}^{3}\Omega/c$ is the total area of the 
polar caps of the neutron star where the roots  of the open 
magnetic field lines are placed, $n_{\pm}$ is the density of the 
particles in the primary beam, $\Omega$ is the angular velocity
of rotation of the pulsar, $R_*$ is the radius of the pulsar
 and the factor $\lambda$ 
takes into account the multiplication of particles because 
the development of electromagnetic cascades in the magnetosphere.
In the  inner gap models,  the
electromagnetic cascade  is initiated by a  beam of 
electrons  accelerated up to the Lorentz-factor 
$\gamma_{\rm gap}\sim 2\cdot 10^{7}$ \cite{rs,arons}.
The density of particles in the
beam is of the order of Goldreich-Julian density 
$n_{\pm}=n_{\rm GJ}$ \cite{gj} determined as
\begin{equation}
n_{GJ}={({\bf \Omega B})\over 2\pi e c}.
\end{equation}
Owing  to the electromagnetic cascade in the pulsar
magnetosphere,  the number of particles increases by a 
factor of $\lambda$ and, correspondingly
the Lorentz-factor of particles decreases by the same factor.

The calculations by Daugherty \& Harding \shortcite{harding} 
and Gurevich \& Istomin \shortcite{istomin} show that for the Crab pulsar 
this factor  is of order of $10^4$.
Note that these early calculations took into account
only multiplication resulting from  the cascades supported  
by two processes -- the curvature radiation of electrons  
and $e^\pm$ pair production of $\gamma$-rays
of these photons in  the magnetic field. 
Meanwhile, the process of the Compton scattering of electrons on 
the soft thermal emission of the neutron star plays, most probably,
a non-negligible role in the cascade development \cite{arons97},
thus this effect should be taken into account in the  estimates of $\lambda$.
In any case,  the uncertainties in the model parameters and
assumptions does not allow an accurate theoretical estimate of 
$\lambda$, but instead  give a broad range of possible values of
$\lambda$ between $10^3$ and $10^5$. Remarkably, a rather 
accurate estimate of $<\gamma_{\rm w}>$
in the upstream flow can be derived from the analysis of the 
nonthermal high energy radiation of the downstream region.
Indeed, the explanation of the spectrum of 
synchrotron X-ray emission by the wind electrons accelerated
(redistributed) by the termination shock requires 
a power-law injection spectrum of the electrons 
$Q(E) \propto E^{-\alpha}$ with $\alpha \simeq 2.4$ and a 
cutoff  of the spectrum below 
$E_{\ast} \sim 150-200$ GeV.
Also, the interpretation of the X-ray and TeV $\gamma$-ray
emissions within the synchro-Compton model of the Crab Nebula
allows one to derive, with very good accuracy, the average magnetic
field in the downstream region and the total luminosity in 
shock-accelerated electrons, $B \simeq 2 \cdot 10^{-4}$~G,
and $\dot W \simeq 3 \times 10^{38} \, \rm erg/s$
(for review see Aharonian \& Atoyan \shortcite{aharonian}). 
The obvious conservation laws
concerning both the number and the total energy of the
of relativistic electrons 
in the downstream and upstream regions (i.e. before and after the
shock acceleration) gives 
$<\gamma_{\rm w}>=\frac{\alpha - 1}{\alpha - 2} \, (E_\ast /m_{\rm e} c^2)
\simeq 1.3 \cdot 10^6$. The accuracy of this estimate 
depends on the possible range of variation of the 
parameters $\alpha$ and $E_{\ast}$ that  still fit the data,
and is estimated  smaller than  factor of 4. Correspondingly
we find the injection rate of the wind particles 
$\dot N = \dot W/<\gamma_{\rm w}> m_{\rm e} c^2 =2.8 \cdot 10^{38}
\, \rm s^{-1}$. This implies that for $\dot N_{\rm gap}= 2\cdot 10^{34}$
the $\lambda$-factor should be close to $10^4$.

Thus, the cascade multiplication of the primary electrons 
accelerated in the pulsar magnetosphere leads to the
formation of an $e^\pm$ plasma
with parameters
\begin{equation}
\dot N= {\lambda B_{0}\Omega^{2}R_{*}^{3}\over ec^{2}},
\end{equation}
and
\begin{equation}
\gamma_{\rm w0}={\gamma_{\rm gap}\over \lambda}.
\label{g0}
\end{equation}
The Lorentz factor of this initial wind is close to
$\gamma_{w0}\sim 10^{3}$ while 
the kinetic energy flux of the wind is
\begin{equation}
\dot E_{0} \approx \gamma_{wo} mc^{2}\dot N
\end{equation}
For the Crab pulsar this flux is estimated as 
$1.6\times 10^{35}$ erg/s.
It is much lower than the total rotational losses of 
the pulsar $\dot E_{rot}=4\times 10^{38}$ erg/s 
(see e.g. Shklovsky \shortcite{shklovskii}). This leads to a
conclusion that the most part of pulsar's rotational 
energy is  carried off by the electromagnetic field (see e.g. Arons \shortcite{arons95}).
The state of the wind is characterized by so-called $\sigma$-parameter
determined as the
ratio of  the electromagnetic energy flux to the kinetic energy flux of
particles in the wind.  At $\sigma_w \ge 1$ the wind is  Poynting-flux
dominated.  At $\sigma_w \le 1$ the wind is  kinetic-energy-dominated.
Initially the wind ejected from the pulsar magnetosphere is 
Poynting flux dominated, because
 $\sigma_{\rm w0}= \dot E_{\rm rot}/\dot E_{0} \sim 2.5\cdot 10^{3}$.
The estimates of the wind parameters in the outer gap model gives
essentially the same magnetization parameter \cite{cheng,kennel90}.

On the other hand, the explanation of the characteristics of the 
nonthermal radiation of the Crab Nebula requires that 
$\sigma_w$ be between $10^{-3}$ and $10^{-2}$
in the wind region upstream of the termination shock 
\cite{rees,kennel}. Thus
the magnetization parameter $\sigma_w$  decreases by several orders of
magnitude on the way from the light cylinder  to the pre-shock region. 
Therefore it is difficult to avoid a conclusion that
the wind is additionally accelerated in a some region beyond the light
cylinder.

The theory by Kennel \& Coroniti \shortcite{kennel} is based on the assumption
of ideal MHD flow of the
plasma after the terminating shock wave. All  dissipative processes are
neglected,  with
exception of the cooling of the plasma due to the synchrotron emission.
Lyubarskii \shortcite{lubarskii1} and Begelman \shortcite{begelman98} argued 
that the theory of Kennel
\& Coroniti \shortcite{kennel} is likely not complete. The magnetic field
after
the terminating shock is mainly toroidal. It is very well
known in plasma physics that such configuration is strongly unstable
\cite{bateman}.
The instabilities can essentially change the
physics of the flow in the Nebula. The basic process is the 
fast dissipation
of the toroidal magnetic field with conversion of its energy into
the energy of the relativistic particles.  The instability drives the plasma 
towards equipartition of energy between 
 the magnetic field and the matter. This
process  will  inevitably be accompanied by acceleration of
particles.
The dissipative processes  provide the dynamics of the
plasma, in good agreement with the observed velocity 
of the expansion
of the outer edge of the nebula,  even for the parameter $\sigma_{\rm w}\sim 1$.
We emphasize that,  although very reasonable,  this argument nevertheless
does not solve the problem of the wind acceleration, because even in this
case we have to transform wind with $\sigma_{\rm w0}\sim 10^{3}$ into wind
with $\sigma_{\rm w}\sim 1$. Moreover, the wind  with $\sigma_{\rm w}\sim 1$ has
almost the same characteristics as we presented above.
The only exception
is that the Lorentz factor of the wind with 
$\sigma_{\rm w}\sim 1$ is twice as small as that
of the wind with $\sigma_{\rm w}\sim 10^{-3}$. In the limits of
uncertainty of the multiplication factor $\lambda$ this 
difference is not important, however

\subsection{The energy spectrum of the wind electrons}

 Here we assume that the wind with $\sigma \ll 1$ is formed
in some `acceleration region' at a distance to the 
axis of rotation   $R_{w}$.  We also assume
that the plasma axially isotropically fills all the open field lines.
 Thus, the wind is not modulated in the azimuthal direction.

For  calculation of the spectra of emission we need  information about spatial
and energetic distribution of  particles of the wind. 
Let us first  summarize the information about these characteristics following directly from 
observations.
The average Lorentz-factor of the wind in the regime
of $\sigma \le 1$ is determined as
\begin{equation}
<\gamma_{\rm w}>={\dot E_{\rm rot}\over \dot N},
\end{equation}
with a typical value $\sim 10^6$.   

The kinetic-energy-dominated wind is believed to be 
cold, because  the region of the
flow  of this wind is observed as an `underluminous'  
region (see Kennel \& Coroniti \shortcite{kennel}
and referenced literature).  Otherwise, a hot wind would produce
remarkable synchrotron emission,  which would contradict  the existence of
underluminous region. 

There is definite latitudinal 
inhomogeneity of the wind. The observations by ROSAT of the torus of X-ray 
emission in the Crab Nebula clearly demonstrate that   
energy flux in the wind varies with latitude \cite{hester}. This fact means that the
density of the wind, its Lorentz-factor and the toroidal
magnetic field should depend on the latitude. 

It is believed that the existence of an X-ray torus implies that
the wind flows predominantly in narrow disc-like sector 
 along the equator with opening angle less than
$30^0$ \cite{hester}. However,  one should take into account 
the fact that optical emission 
is uniformly distributed in the Nebula. This means  that there should be a component
of the  wind at high
latitudes.  As  the luminosities in optical and in X-rays are comparable 
the energetics of this  high latitude wind component is  
 comparable with the energetics of the equatorial component which produces 
the X-ray torus.
Moreover, the evidence that the wind at high
latitudes exists close to the pulsar follows also from HST observations \cite{hester}. 
In our  calculations we assume that the particle flux in the wind is 
isotropic. However to be consistent with observations in X-rays we 
assume that the Lorentz-factor  and the toroidal magnetic field 
of the  wind depend on latitude. 

There are no model independent estimates of the latitudinal dependence of the 
Lorentz-factor of the wind and toroidal magnetic field.
We use here the characteristics of the wind  obtained in the model of an axisymmetric 
rotator \cite{bog97}. These characteristics are also valid for
oblique rotators  \cite{bog99}.

Below we assume that
\begin{equation}
\gamma_{\rm w}=\gamma_{\rm w0} + \gamma_{\rm max}\cos^2(\alpha),
\label{9}
\end{equation}
 where $\gamma_{\rm max}=\sigma_{\rm w0}\gamma_0$ is the maximum Lorentz-factor of the
plasma on the equator($\alpha=0$); $\gamma_0 \sim 10^3$ (see equation \ref{g0})
and $\sigma_{w0} \sim 10^3 $  are the  
Lorentz-factor and the magnetization parameter of the wind near the light cylinder.
The wind is assumed to be monoenergetic at a given  latitude. 
It follows from equation (\ref{9}) that
\begin{equation}
<\gamma_w>={2\over 3} \gamma_{max} \, .
\end{equation}  
In the particular case of the axisymmetric rotator \cite{bogkot}, 
\begin{equation}
\gamma_{max} = {eB_{0}R_{*}\over 2mc^{2}\lambda}(R_{*}\Omega/c)^{2}.
\end{equation}

\subsection{The geometry of the wind flow}

\begin{figure}
\epsfxsize=8.0 cm
\centerline{\epsffile[0 0 572 574]{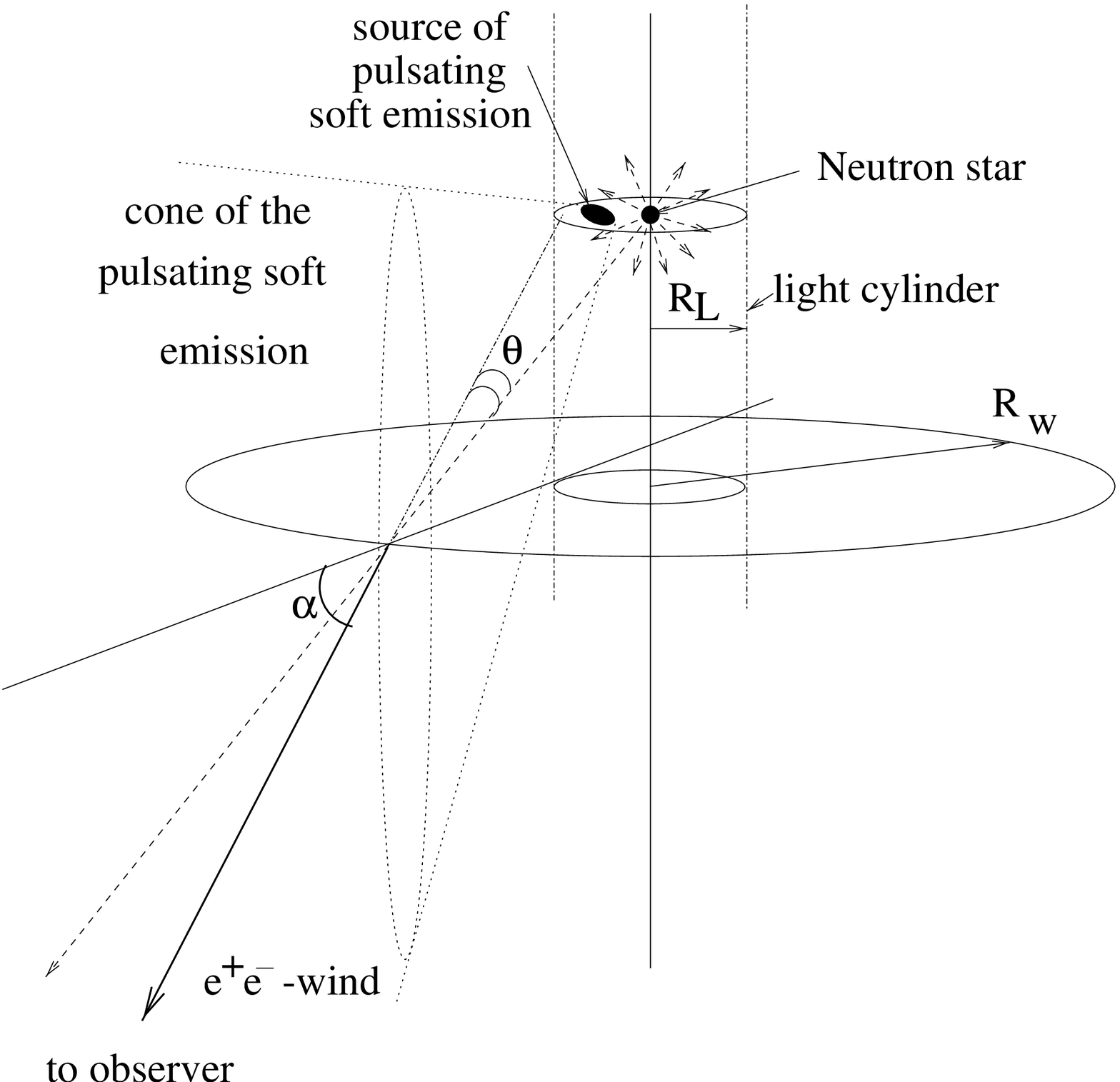}}
\caption{A sketch of the trajectories of plasma after
acceleration, and the assumed position of sources of soft photons.}
\label{fig1}
\end{figure}

The geometry of the flow and of the IC process are  drawn in fig. \ref{fig1}.
The neutron star is placed on the axis of rotation and ejects the wind radially. 
The dash-dotted vertical lines show the 
light cylinder. The wind is accelerated in the acceleration zone by some 
unspecified mechanism and at the distance $R_W$ it has characteristics discussed
above. The acceleration is completed at $R_W$. Beyond $R_W$ particles in the wind move
along straight lines without further acceleration.
The equatorial  plane of the pulsar is inclined in relation to the 
observer at the angle $\alpha = 33^0$  \cite{hester}.
IC photons move along the  direction of motion of electrons. Therefore only the 
particles of the wind directed towards   Earth can   produce observable emission. 
The lines of
flow of the cold kinetic energy dominated wind after acceleration are not 
exactly radial. 
Classical mechanics provides a simple relationship between the 
rates of the rotational energy losses of the pulsar, 
$\dot E_{rot}=I\Omega\dot \Omega$, and the 
angular momentum losses, $\dot L=I\dot\Omega$: 
\begin{equation}
\dot E_{\rm rot} = \dot L\Omega.
\label{eq0}
\end{equation}
Here $I$ is the momentum of inertia  of the neutron star. 
As  the electromagnetic field carries
off practically no
energy in the kinetic-energy-dominated  wind, 
it does not carry the  angular momentum either. In the
theory of plasma flow in the magnetosphere of an axisymmetric rotator this statement
is verified immediately \cite{bog97}. Under this condition it follows from
equations (\ref{eq1}),(\ref{eq2}), and (\ref{eq0}) that after the acceleration the azimuthal
velocity of the wind $v_{\varphi}$ is connected with $R_W$ as 
\begin{equation}
  {v_{\varphi}\over c}={R_{\rm L}\over R_{\rm W}}.
  \label{vf}
\end{equation}
From this relationship and from Fig. \ref{fig1} it follows that
the projection of the vector of the velocity of the plasma after acceleration 
lies on the line tangential to the light cylinder.
The angle $\theta$  between the direction of the motion of relativistic 
particles in the wind and the soft thermal photons emitted from the  pulsar  depends on
distance $r$ to the axis of rotation:  
\begin{equation}
\sin{\theta}=\cos{\alpha}{R_{\rm L}\over r}.
\label{sin}
\end{equation}
It is seen from this relationship that at any distance from the pulsar
there is a non-zero angle
between the radial direction and the velocity of the wind.  
The Inverse Compton scattering of the 
wind electrons on  soft photons emitted as a fan-like beam  from the inner
magnetosphere of the
pulsar results in  the production of hard $\gamma$- ray photons.
Equations (\ref{sin}) and (\ref{vf}) remain valid for condition
$\gamma_{w} \gg \gamma_{w0}$ \cite{bog97}, also  fulfilled in the wind with
$\sigma_{w}\sim 1$. Below we will not distinguish between the cases with
$\sigma_w \ll 1$ and $\sigma_{w} \sim 1$ .

The angular distribution of soft emission close to the pulsar is not well known.
In the outer gap model the  optical and soft x-ray emission is generated inside, but not
 far from the light cylinder \cite{romani}. There are 
several sources of soft emission in the  magnetosphere corresponding to the position of the 
outer gaps. In this model,  the projection of  the  motion
of photons from these sources on the equatorial plane 
is predominantly directed radially. Then the angle between  direction of motion 
of the wind and soft non-thermal photons can also be estimated by equation (\ref{sin}). 
Note  that at $R_W \gg R_L$ the IC flux does not  depend on the position of the source 
inside the magnetosphere if the soft photons are emitted radially. 

Our calculations of IC radiation  are based on the assumption that the wind is 
illuminated only by the emission that is
directed to Earth. However, as  we can not exclude the existence of an additional component
of optical emission not directed to the observer, but illuminating the wind, our estimates of
$\gamma$ -ray flux could be considered only as a lower limit. 

\section{Inverse Compton radiation of the wind}

\subsection{The fields of soft photons}

The soft emission from the Crab pulsar consists of two, pulsed 
and unpulsed,  components. The recent ROSAT 
observations  allowed one to distinguish  the contribution of the pulsar
in the observed unpulsed flux from the emission of the  nebula. 
The main contribution to the unpulsed radiation of the pulsar comes, most probably,
from the thermal emission of
the hot surface of the neutron star.   According
to the ROSAT observations,  the total energy flux of the unpulsed  emission 
in the range 0.1 - 2.4 keV is $10^{34}$ erg/s \cite{trumper}. 
The latter could be approximated by 
black-body spectrum with a 
temperature $1.9\cdot 10^{6} K$ 
and  total luminosity $10^{34}$ erg/s.

The pulsed soft radiation of the pulsar is dominated
by nonthermal processes  in the 
magnetosphere. The photons in the optical to X-ray band
of this radiation play the most important role in
the production of inverse Compton $\gamma$-rays.
Measurements of the spectra of the pulsed optical emission 
by different groups \cite{nasuti,oke,percival} give
the following spectra of  optical photons in the range 1680 - 7400 {\AA}
\begin{equation}
F_{\nu}=3.1 ({\nu\over \nu_{0}})^{-0.11} \, \rm mJy,
\end{equation}
where $\nu_{0}= 6.82 \cdot 10^{14} \, \rm Hz$.
 
The X-ray data in the range of 0.1 - 2.4 keV obtained 
by ROSAT can be approximated by a power-law  with the 
photon index 1.5 and luminosity (assuming that the
radiation is emitted isotropically) $7.1 \cdot 10^{35}$ erg/s
\cite{trumper}.

\begin{figure}
\epsfxsize= 8.0cm
\centerline{\epsffile[89 223 462 585]{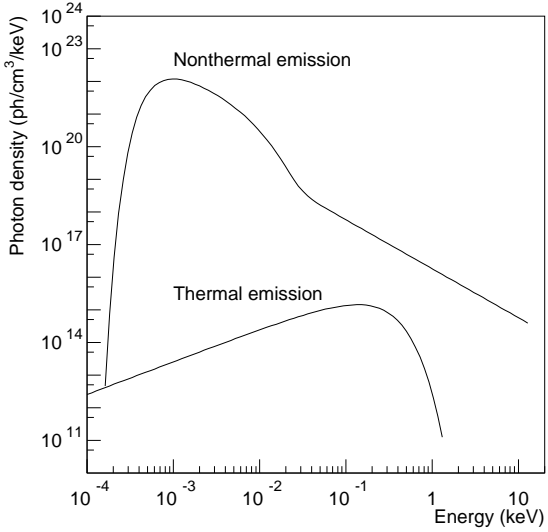}}
\caption{The averaged spectral density of photons near the 
light cylinder of the Crab
pulsar}
\label{fig2}
\end{figure}

The extrapolation of the soft x-ray emission 
spectrum to the optical range gives lower flux than is observed  
\cite{knight}. This means that the optical emission has a cutoff in the
ultraviolet region. Unfortunately there are no data for  the 
pulsating emission in this region. Therefore we assume an exponential 
cutoff in the spectrum above 0.05 keV.
The emission is also strongly suppressed at infrared wavelengths 
\cite{middleditch}.
Below we use the following approximation of the averaged  density of
non-thermal photons  near the light cylinder
\begin{equation}
n(\epsilon) =Z(\epsilon) \, \exp{\left[-\left({8\cdot 10^{-4} 
~{\rm keV}\over\epsilon}\right)^{2} \right]}\cdot 
 {\rm photon\over cm^{3} keV},
\end{equation}
where $\epsilon$ is the energy of photons in keV and
\begin{eqnarray}
\lefteqn{Z(\epsilon)=1.82\cdot 10^{16} \epsilon^{-1.5}+}\nonumber\\
& & +1.25\cdot 10^{19}\epsilon^{-1.11}
\exp{\left[- \left(\epsilon\over
5\cdot 10^{-3} ~{\rm keV} \right) \right]}
{\rm ph\over cm^{3} {\rm keV}}.
\end{eqnarray}

The photon  density  of thermal and non-thermal components of 
low-frequency radiation on the light cylinder
is shown in Fig. \ref{fig2}.
We assume that the density of the both photon fields fall off 
with distance to the pulsar
center $R$ as  ${1/ R^{2}}$.

\subsection{The fluxes and spectra of gamma radiation}

The optical depth characterizing the Compton scattering of 
the wind electrons propagating through the radiation fields
of the pulsar is defined as   
\begin{equation}
\tau = \int \sigma (\omega, \gamma, {\bf v}, {\bf c})(1-{\bf vc})
n_{ph}(\omega, {\bf c}, r)dl d\omega,
\label{opd}
\end{equation}
where $\sigma (\omega, \gamma, {\bf v}, {\bf c})$ is the 
total invariant cross-section of the IC scattering, 
$\omega$ is the energy of soft photons in $mc^{2}$.  
Integration of equation (\ref{opd}) over $dl$ is performed along 
the trajectory of a wind electron (see  Fig. 1) from the wind formation
position to infinity.

The $\gamma$-ray energy flux at the angle  $\alpha$ to the plane of the  
pulsar equator   is equal to
\begin{eqnarray}
\lefteqn{L_{\rm IC}(E_{\gamma})= {\dot N\over 4\pi D^2}\times} \nonumber\\
& & \times\int \int E_{\gamma} {d\sigma\over dE_{\gamma}}
(\gamma_{\rm w}(\alpha),\omega, E_{\gamma},\theta) n_{ph}(\omega, {\bf c}, {\bf r})d\omega dl, 
\label{sp}
\end{eqnarray}
where $D=2$ kpc is the distance to the source, ${d\sigma\over dE_{\gamma}}
(\gamma_{\rm w}(\xi),\omega, E_{\gamma},\theta)$ is the differential cross-section of the 
Compton scattering of a photon with  energy $\omega$ and electron with 
$\gamma_w$ encountering under angle $\theta$ and producing a photon with  energy $E_{\gamma}$.
This cross-section was obtained after  integration of the differential cross-section \cite{landau} 
over the emission angle
\begin{eqnarray}
{d\sigma\over dE_{\gamma}}
(\gamma,\omega, E_{\gamma},\theta)={\pi r_{\rm e}^2\over (pk)^2}\{{1\over (pk)^2}-
{2I\over (pk)\gamma_c\omega_c}+\nonumber\\
+{I^3(1-U\delta\beta)\over (\gamma_c\omega_c)^2}
+2({1\over (pk)}-{I\over \gamma_c\omega_c})
+\nonumber\\
+({(pk)I\over \gamma_c\omega_c}+{\gamma_c\omega_c(1-U\delta\beta)\over (pk)})\}
{\omega_c\over \Gamma V_c},
\label{19}
\end{eqnarray}
where
\begin{eqnarray} 
(pk)=\gamma\omega(1-v\cos{\theta}), & V_c={\sqrt{\omega^2+(\gamma v)^2
+ 2\gamma v \omega\cos(\theta)}\over (\gamma+\omega)} &,\nonumber\\
\Gamma={\gamma+\omega\over \sqrt{1+2(pk)}}, &
\omega_c={(pk)\over  \sqrt{1+2(pk)}},\nonumber\\
  \gamma_c={1+(pk)\over \sqrt{1+2(pk)}}, & U = {(pk)\over 1+(pk)},\nonumber\\
\delta ={1\over V_c}({E_{\gamma}\over \Gamma\omega_c}-1),& 
\beta ={1\over UV_c}({\gamma\over \Gamma\gamma_c}-1).\nonumber
\end{eqnarray} 
Function $I$ in equation (\ref{19}) is defined as follows
\begin{equation}
I=(1-U^2-2U(1-U)\beta\delta+U^2(\delta-\beta)^2)^{-1/2}. 
\end{equation}

Integration of equation (\ref{sp}) is  performed  along the trajectory of a electron 
and over the spectra of soft photons.
It is assumed   that the observer detects  photons from a monoenergetic beam of electrons 
from  the wind moving  to the observer.
We take $\gamma_{\rm max} ={3\over 2}\dot E_{\rm rot}/\dot N$, 
with $\gamma_{\rm w}(\alpha)=\gamma_{\rm max}\cos^2(33^0)$, since 
the plane of the pulsar equator is inclined to the observer at  the angle $33^0$ \cite{hester}.

\subsection{IC photons from thermal isotropic radiation}

The optical depth $\tau$ , and therefore the 
spectra of IC photons, depends strongly on the 
distance $R_{\rm w}$ at which  the kinetic energy dominated wind is formed.

\begin{figure}
\epsfxsize= 8. cm
\centerline{\epsffile[89 223 462 585]{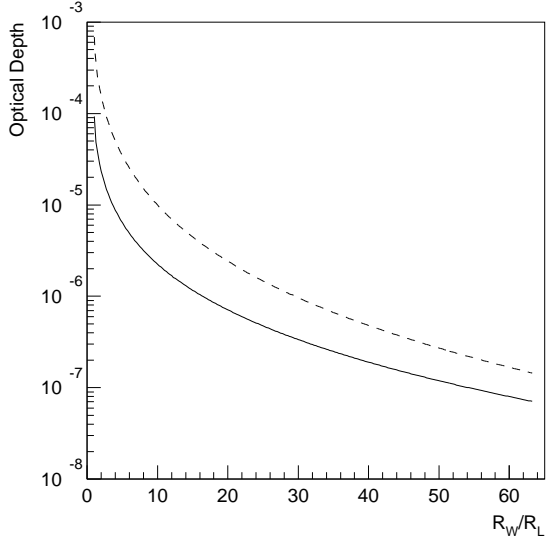}}
\caption{The  optical depth  of the thermal photon field 
for  the IC scattering of  electrons
with Lorentz factor $\gamma_{\rm w}=4\cdot 10^{6}$ (dashed line) and  
$4\cdot 10^{7}$ (solid line).}
\label{fig6}
\end{figure}

The dependence of the optical depth on $R_{\rm w}$ 
is shown in Fig. \ref{fig6}  for two different values of the 
Lorentz factors of the wind $\gamma_{\rm w}=4\cdot 10^{6}$ and
$4\cdot 10^{7}$
%
Strong dependence of $\tau$ on $\gamma_{\rm w}$ 
is explained by the  fact that  the optical depth is 
dominated  by IC scattering at 
small  distances from the pulsar where 
$2\omega\gamma(1-\cos{(\theta)}) \gg 1$, i.e. 
the Compton scattering takes place in the Klein-Nishina regime.
In this regime the IC cross-section decreases with the electron
energy as $\gamma_w^{-1}$, which lead to larger optical depth for smaller values
of the Lorentz-factor of the wind $\gamma_{w}$. This effect is especially
strong for small values of $R_{\rm w}$. In this case the optical
depth is accumulated by IC scattering close to the light cylinder
where the collision angle $\theta$ is large, and the IC scattering takes place
in deep Klein-Nishina regime.

In many ``standard'' 
astrophysical situations the efficiency of the 
$\gamma$-ray production in the Klein-Nishina regime
is significantly suppressed because of the synchrotron cooling of  
electrons. In the case of a {\it cold} wind we deal with 
a unique situation when the  synchrotron losses 
of electrons are completely suppressed, and thus the
wind electrons lose their energy only through the inverse
Compton radiation. Note
that in  deep Klein-Nishina regime, namely
when $2\gamma_w \omega (1-\cos\theta) \geq 10^4$, the triplet pair
production dominates over the inverse Compton scattering
\cite{mastichiadis,dermer}. This process results in production of new 
electrons. However, as  in all interesting cases the optical depth 
$\tau \le 1$, the secondary electrons do not increase significantly the
density of the wind. Therefore we ignore this process here.
The same is true also for another process connected with absorption of
TeV $\gamma$ -rays in the magnetic field of the wind (see Appendix).

In fig. \ref{fig8} we present the 
expected $\gamma$-ray fluxes of the wind. 
%
\begin{figure}
\epsfxsize= 8. cm
\centerline{\epsffile[89 223 462 585]{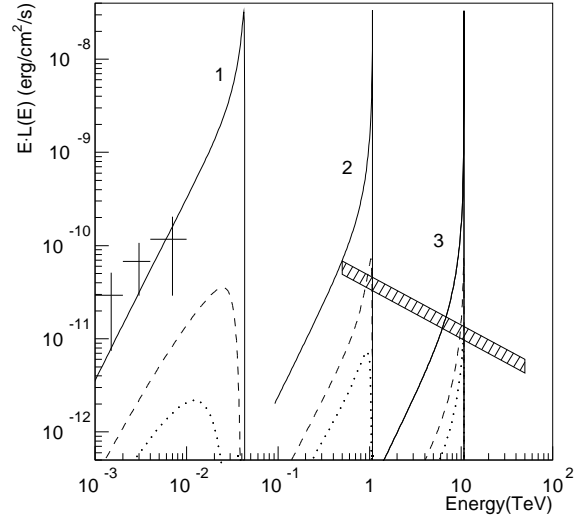}}
\caption{The spectra of IC radiation of the wind
illuminated by the thermal emission of the pulsar. It is assumed that
the kinetic-energy-dominated wind is formed at distances of 1 (solid),
5 (dashed), and 10 (dotted) light cylinders from the pulsar. The  curves  
1,2,3  correspond  to
$\gamma_{max}=1.2\cdot 10^5, 3\cdot 10^6, 3\cdot 10^7$. The  range of observed 
fluxes of $\gamma$-rays in the  region above 500 GeV
detected by CAT, CANGAROO, HEGRA and Whipple telescopes
is  shown by a shadowed region. The points with error bars
below 10 GeV correspond to the unpulsed fluxes measured by 
EGRET.}
\label{fig8}
\end{figure}
%
\begin{figure}
\epsfxsize= 8. cm
\centerline{\epsffile[89 223 462 585]{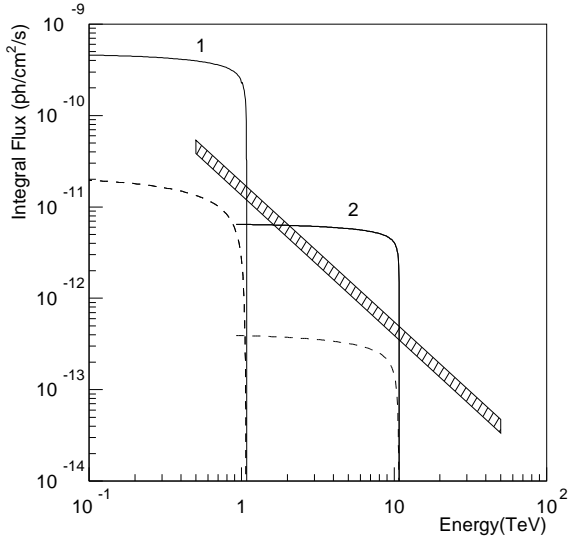}}
\caption{The integral fluxes of IC radiation of the wind
illuminated by the thermal emission of the pulsar.
It is assumed that the kinetic-energy-dominated wind is formed at 
distances of 1 (solid line ) and 5 (dashed line ) light cylinders from the pulsar.   
The curves  1 and 2  correspond  to
$\gamma_{max}=3\cdot 10^6$ and $ 3\cdot 10^7$ respectively. The  range of observed 
fluxes of $\gamma$-rays in the  region above 500 GeV
is  shown by a shadowed region.} 
\label{fig8a}
\end{figure}
Solid, dashed and dotted lines correspond to the fluxes produced by a wind
originated at $R_{\rm w}/R_{\rm L}$=1, 5 and 10. 
Remarkably, the spectrum of IC $\gamma$ -rays has a specific  
line feature since the radiation is produced in  the Klein-Nishina regime.
Because of this feature, the IC radiation of the wind can be 
 easily distinguished from the
smooth spectra of the Crab Nebula. 
The corresponding integral fluxes are shown in Fig. \ref{fig8a}
for different $\gamma_w$ and $R_w$.

The comparison of the calculated  spectra  with the observed  TeV
$\gamma$-ray fluxes of the Crab Nebula (for review see Weekes  et al. 1997)
leads to an interesting conclusion 
that  for any reasonable Lorentz-factor of
the wind ,
the calculated $\gamma$-ray fluxes significantly  exceed the observed  
fluxes unless the wind is  formed well beyond the light cylinder.
This implies  a meaningful constraint on the `birthplace' of the
kinetic energy dominated wind with 
$\gamma_w \ge 10^6$: $R_{\rm w} \geq 5 R_{\rm L}$.
For $\gamma \le 1.2\cdot 10^5$,  a similar conclusion 
is imposed by the EGRET \cite{nolan} 
observations of unpulsed radiation above 1 GEV.
The lack of measurements in the energy region 
between 10 GeV and 300 GeV does not completely 
exclude  a possibility of formation of the particle dominated wind 
close to the light cylinder, if one 
assumes that  $10^5 < \gamma_{max} <  10^6$. Although the 
analysis of the  observed synchrotron and IC components of
the nonthermal radiation of the Crab Nebula gives certain  
preference to larger wind  Lorentz factors, $\gamma_{\rm w} \sim 10^6$, 
it is important to have  {\it observational} constraints 
on $R_{\rm w}$ also for $\gamma_{\rm w}$ in the  
region between $10^5$ to $ 10^6$, which is presently missing This should be 
provided  in the near future  by 
measurements of the Crab spectrum between  10 and 300 GeV by 
the new generation of low-threshold atmospheric Cherenkov detectors
as well as by the GLAST.  

\subsection{The interaction of the wind with the nonthermal emission}

\begin{figure}
\epsfxsize= 8. cm
\centerline{\epsffile[0 0 449 403]{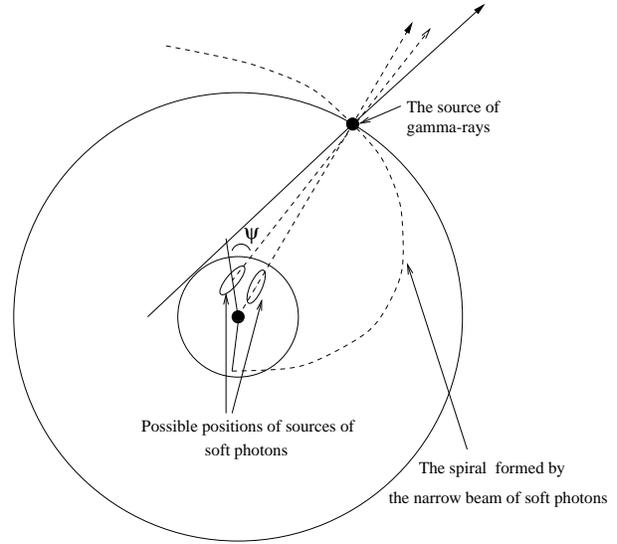}}
\caption{The angle between the wavevector of soft 
photon and the velocity of particles
in the wind depends on the position of the 
source of soft photons inside the magnetosphere
and the direction of the beam. 
The minimum efficiency of the gamma-ray production
is achieved when the source is corotating with  
the pulsar. The narrow beam of
non-thermal soft photons produces a spiral in  space. 
The point of crossing of this spiral with
the trajectory of an electron moving to an observer is the source of 
$\gamma$-ray  photons,  which moves with a velocity
very close to the light velocity. At small angles between 
the beam of the soft photons
and the velocity of particles in the wind, the delay 
in the arrival time of hard photons and optical photons
is small. The width of the gamma-ray pulse is also small. 
The phase curves for  optical radiation 
illuminating the wind and in VHE gamma-rays
are  expected to be similar.}
\label{fig11}
\end{figure}

The IC $\gamma$ radiation caused by illumination of the wind by pulsed soft photon emission 
should be modulated at the same pulsation period. 
 This conclusion is true even for the
isotropic wind. For calculation  of the IC fluxes it is assumed that the nonthermal source of 
soft emission is located inside the light cylinder  (see Fig. \ref{fig1}).
It is easy to show that under realistic conditions the phase curve of the
gamma-rays should be  close to the phase curve  of the  soft
emission. 
Indeed, let us assume for simplicity that the beam is very narrow (delta-function like). 
Owing to the rotation of the pulsar,  it will form
a spiral. The point where this spiral crosses the line  tangential to the light cylinder and
directed towards the observer is the source of $\gamma$-rays (see Fig. \ref{fig11}). 
The width of the $\gamma$-ray pulse is determined by the delay in 
the arrival time $\delta t$ of optical and gamma-ray emission to the observer:
\begin{equation}
\delta t \sim {\theta T\over 2\pi },
\end{equation}
where $\theta$  is the maximal value of the angle between the wave vector of photons and the
velocity of the particles in the wind, $T$ is the period of rotation. 
In the most interesting case,  $\theta \ll 1$, the phase curves of gamma-ray
emission produced in the process under consideration and those 
of the optical emission illuminating the wind are expected 
to be similar. However,  because we 
observe only the pulsed soft emission directed to us and do not observe other 
possible part of the emission (not directed towards the observer) 
that can illuminate the wind, there could be a difference 
between the $\gamma$-ray and directly observed soft emission light curves.

\begin{figure}
\epsfxsize= 8. cm
\centerline{\epsffile[89 223 462 585]{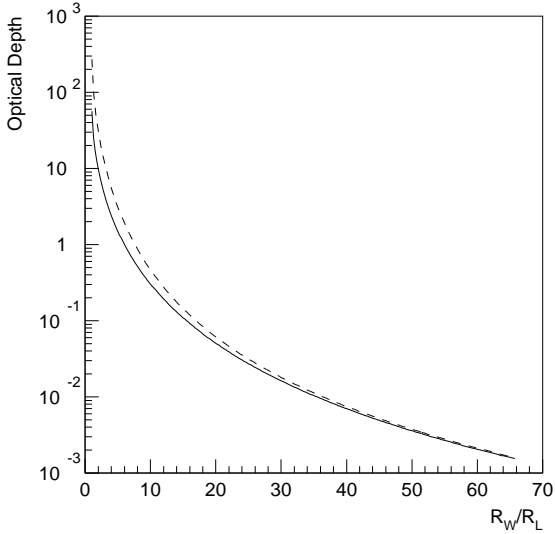}}
\caption{The optical depth   for IC scattering  of electrons on 
nonthermal pulsating radiation of the
pulsar. It is assumed that the nonthermal photons
are directed radially.
Curves are given for $\gamma_{w}=4\cdot 10^{6}$ (dashed line)
and for $\gamma_{w}=4\cdot 10^{7}$ (solid line).}
\label{fig9}
\end{figure}

 \begin{figure}
      \epsfxsize= 8. cm
\centerline{\epsffile[89 223 462 585]{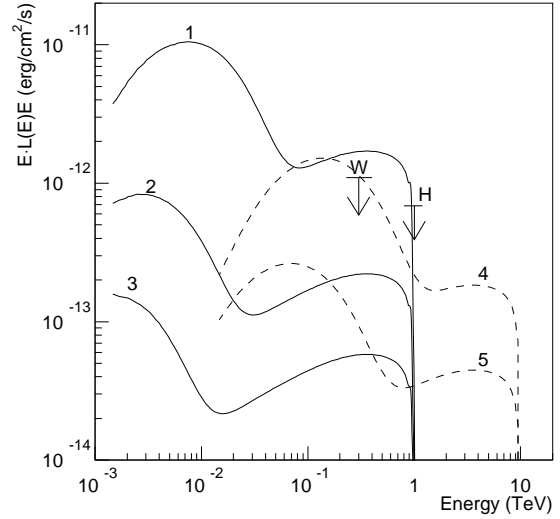}}
\caption{The spectra of pulsating TeV gamma-ray emission produced by the wind with
$\gamma_{max}=3\cdot 10^6$
(solid lines) and   $\gamma_{max}=3\cdot 10^7$ (dashed lines)  
Numbers 1,2 and 3 near the solid lines correspond to 
$R_{w}/R_{L}=30, ~50$ and  70,  respectively. The curves 4 and 5 correspond to 
$R_{w}/R_{L}=70$ and  100.
The upper limits
on the pulsed  radiation from Crab Nebula  above 300 GeV given by
Whipple group \protect\cite{weekes} and above 1 TeV given 
by HEGRA collaboration \protect\cite{aharonian98}
are also shown. The beam  of  soft photons is assumed to be radial.}
\label{fig13}
\end{figure}

It is seen from  Fig. \ref{fig2} that the average
density of nonthermal photons is several orders  of 
magnitude larger  than the density of the 
thermal photons. Therefore,  the IC optical depth for an electron 
is much larger  than in the case of thermal radiation (see Fig.\ref{fig9}).
In particular,  at $R_W \le 5R_L$ the optical depth $\tau \ge 1$. 
As  the IC scattering takes
place in the Klein-Nishina regime, one or two interactions 
are sufficient to destroy the wind, and
the whole energy of the wind electrons would be 
transferred to $\gamma$ ray emission with huge
luminosity $\ge 10^{38}$ erg/s. This very fact excludes the possibility of  
formation of the kinetic dominated wind within $5R_L$. Moreover, 
this conclusion can be extended to larger  distances.
Fig.  \ref{fig13} shows the spectra of the emission
for two maximal energies of the electrons in the wind. 
Solid lines show the spectra for  $\gamma_{max}=3\cdot 10^6$ and
dashed lines show the spectra for $\gamma_{max}=3\cdot 10^7$.  
In contrast to the IC $\gamma$-rays from interaction with thermal emission, 
there are no  lines  in the spectra of the emission because  the IC scattering  
takes place in the Thomson regime. 
The spectra consist of two well separated broad components, 
corresponding  to two components
in the spectra of the soft nonthermal emission presented in Fig. \ref{fig2}.
The comparison of calculated $\gamma$-ray fluxes with upper 
limits on pulsed emission  reported by the
Whipple \cite{weekes} and HEGRA \cite{aharonian98} groups shows that  
the wind must be formed at relatively large distances from the pulsar,
$R_W > 30 R_{L}$.

The spectra shown in Fig. \ref{fig13} were calculated at the assumption that the 
nonthermal source emits photons along the radial direction from  the pulsar as it 
happens in the
outer gap model \cite{romani}. Self-consistent MHD solutions confirm that the plasma
should move predominantly  radially in the pulsar magnetosphere \cite{bog97}. Therefore the 
 source of nonthermal photons should indeed emit along the radial direction from the pulsar.
In the model by Lyubarskii \shortcite{lubarskii2} the source  of the soft nonthermal 
photons emits in the direction opposite to the direction of rotation of the pulsar.
If so, the $\gamma$ -ray emission would be even higher than shown in Fig. \ref{fig8}.

We are not aware of models in which the source 
of photons corotates with the pulsar and emits photons  tangentially along the
direction of rotation. In this case the angle of interaction $\theta$ is minimum and 
correspondently the IC flux is expected to be reduced. Although 
it is almost unlikely that the plasma
(and source of soft photons ) 
can corotate with the pulsar, it is perhaps wise 
to consider this limiting case from pedagogical point of view in some detail.

Optical and soft X-ray emission can be produced by  plasma moving with a 
Lorentz-factor of  order $\gamma_e \sim 10^2-10^3$
 in  the pulsar magnetosphere. At  corotation
with the pulsar  the plasma  have   azimuthal velocity 
$v_{\varphi}={r\Omega\over c}$, where $r$ is the 
radius in the cylindrical system of coordinates. Since 
these particles should also be directed towards the observer, they have the component
of the velocity $v_z= c\cos\alpha$. From the relativistic relationship it follows
that
\begin{equation}
\gamma_e^2=1+\gamma_e^2[cos^2\alpha+({r_s\Omega\over c})^2+({v_r\over c})^2],
\end{equation}              
where $r_s$ is the distance from the source to the center of the pulsar, 
$v_r$ is the radial component of the velocity.
The soft photons are emitted
at the angle $\psi$ as shown in Fig. \ref{fig11}.
Neglecting the term $1/\gamma_e^2$ we obtain that 
\begin{equation}
\sin\psi = {r_s\Omega\over c\sin\alpha}.
\end{equation} 
This relationship shows that a  source that emits  soft photons   
 in direction to  the observer and corotating with the pulsar can not 
be located exactly on the light cylinder.

\begin{figure}
\epsfxsize= 8. cm
\centerline{\epsffile[89 223 462 585]{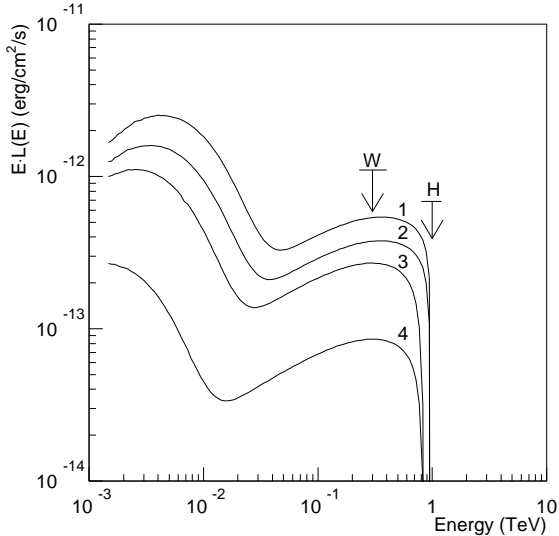}}
\caption{The dependence of the  differential spectra 
of $\gamma$ -ray emission on the position of the source of soft photons 
inside the pulsar magnetosphere
for  parameters of the wind  $\gamma_{max}=3\cdot 10^{6}$ at $R_W=40R_L$. 
The source of soft  photons is placed on the distance 
$r_s=0., 0.3, ~0.5,~0.6R_{L}$ respectively for curves 1, 2, 3, 4. 
The direction of the beam of the soft photons was 
defined under the assumption of corotation
of emitting particles with the pulsar.}
\label{fig12}
\end{figure}

 Fig. \ref{fig12} demonstrates how
the spectra of IC radiation  depend  on the position of the corotating source of the 
soft photons inside the pulsar
magnetosphere for parameters of the kinetic energy dominated wind at  $R_w=40R_L$.
Curve  1 corresponds to the emission from the corotating source located on the surface of 
the pulsar. The same flux is produced by the source of soft photons with the beam directed
radially from the pulsar;  the IC flux generated from the last source does not depend on $r_s$.
Therefore comparison of the curve 1 with others allows one to compare the fluxes of IC photons
produced at the scattering of electrons on the soft emission  
from the corotating source and the source with
the soft photons  emitted radially, but located on the same distance $r_s$.    
It follows from this figure that the flux of the IC photons can be reduced to zero only at
the position of the source of soft photons  not far from  the light cylinder. At any other
position of the source the efficiency of generation remains high  at $r_s < 0.6 R_L$.
The existing upper limit on the  pulsating flux from the Crab pulsar means
that either the wind is formed beyond 30$R_{L}$ or the wind is formed close to the
light cylinder but the pulsating source of soft nonthermal 
photons corotates with the pulsar and  is located   close to the
light cylinder.

\section{Summary}

The Crab Nebula is a unique cosmic laboratory 
with an unprecedentedly broad spectrum of the observed
nonthermal radiation that extends throughout  21  decades of 
frequency -- from radio wavelengths  to very high 
energy $\gamma$-rays (for review see e.g. Aharonian \& Atoyan 1998)
It is commonly accepted that the synchrotron nebula is powered by 
the relativistic wind of electrons generated at the pulsar and
terminated by a standing reverse shock wave at a distance 
$r_{\rm s} \sim 0.1 \, \rm pc$ \cite{rees}.    
The relativistic MHD models,  
even in their simplified form (e.g. ignoring the 
axisymmetric structure  of the wind  and its interaction with the optical 
filaments),  successfully describe 
the general characteristics of the synchrotron nebula, and
predict  realistic distributions of relativistic
electrons and magnetic field 
in the downstream region behind the shock \cite{kennel}.
  Meanwhile  our knowledge of the unshocked wind,
i.e. about the region between the pulsar magnetosphere and the shock is
based only on theoretical speculations. Moreover, it is commonly
thought that the wind could not be visible in the region upstream
of the termination shock because the relativistic electrons and 
magnetic field in  wind move together, thus the unshocked 
wind does not produce synchrotron radiation.      
In this paper we show that the kinetic energy dominated wind 
nevertheless could be directly observed through its IC radiation
because of  the illumination of the wind by low-energy radiation of the
pulsar. The $\gamma$-ray emission  consists of two components, pulsed and unpulsed,
associated with the nonthermal and thermal 
low-energy radiation components of the pulsar, respectively. 

The unpulsed component of $\gamma$-ray emission 
associated with thermal radiation of the pulsar with temperature
$\simeq 2 \cdot  10^6 \, \rm K$  is produced in 
 deep the Klein-Nishina regime, and therefore has a very sharp (line-like) 
spectral feature which peaks at energy $E \simeq \gamma_w \, \rm m_{\rm e} c^2$.
Detection of this component would therefore result in  
 unique information about the Lorentz-factor of the bulk 
motion of the wind.  The nonthermal radiation of the pulsar has rather broad
energy spectrum which extends to optical and infrared wavelengths, 
and therefore the IC $\gamma$-ray emission
associated with this component  takes place, to a large extent,
in the Thomson regime.  This results in a broad $\gamma$-ray spectrum
with a sharp cutoff at  $E \simeq \gamma_w \, \rm m_{\rm e} c^2$.

The absolute $\gamma$-ray fluxes of both components depend
strongly on the site of formation of the kinetic dominated wind,
as well as  the Lorentz-factor and the geometry of propagation 
of the wind.  Thus even the flux upper limits of these 
$\gamma$-ray components should provide important constraints
on the wind parameters. In particular, we show that the comparison of the calculated flux  of
the unpulsed inverse IC emission with the measured $\gamma$-ray flux of
the Crab Nebula excludes the possibility of formation  of the 
kinetic-energy-dominated wind within
5 light cylinder radii of the pulsar, $R_{\rm w} \geq 5 R_{\rm L}$.
 The analysis of  the pulsed IC emission,
calculated under reasonable assumptions concerning the production site and
angular distribution of the optical pulsed radiation,
yields even tighter restrictions, namely $R_{\rm w} \geq 30 R_ {\rm L}$.

The mechanism of $\gamma$-radiation of the wind of the 
Crab pulsar discussed in this paper should certainly 
take place in other pulsars as well.  However, from the point of
view of detection of this radiation, the Crab is a unique object
due to its very powerful wind and relatively high 
luminosity of thermal and nonthermal low-energy radiation, 
which provides seed photons for the IC scattering.  In other 
pulsars the IC $\gamma$-ray fluxes of unshocked winds 
are expected to be below the detection threshold of current 
$\gamma$-ray instruments, unless the kinetic energy dominated winds 
of pulsars are  produced very close to the light cylinder. 
The situation could be different in binary systems containing 
a pulsar and luminous optical companion, 
the latter being an effective supplier of seed photons for
IC scattering. For example, the  pulsar/Be star binary system 
PSR 1259-63 seems to be a unique object for the search for IC TeV 
radiation from both shocked \cite{kirk1} and unshocked 
\cite{kirk2}  winds of the pulsar.  

\noindent
{\bf Acknowledgments} 

We thank the Astrophysics group of the 
MPI f\"ur Kernphysik, in particular,  H. J. V\"olk,  
A.M. Atoyan,  J.G. Kirk,  as well
as L. Ball and Yu. Lyubarskii  for many 
fruitful  discussions.  SB thanks  MPI f\"ur Kernphysik
for warm hospitality and support during his work on this
project.

\appendix
\section{Opacity of the magnetic field of the wind for the TeV Photons.}

In the paper of \cite{cheng} it was argued that TeV photons produced near the light cylinder
of the Crab pulsar will be absorbed,  because of conversion of these photons in
$e^{\pm}$ pairs in nonuniform  magnetic field. We show here that accurate estimate
of the rate of the conversion of photons in pairs taking into account electric field existing in the 
wind give negligible  absorption of TeV gamma-rays produced via IC scattering.

The motion of the wind occurs under frozen in
condition. Particles move not only along field lines, but there is component of the
velocity ${\bf V_{d}}$ directed perpendicular to the magnetic field line. This is so called
drift velocity defined by the expression
\begin{equation}
 {\bf V_{d}}={ {\bf E\times B}\over B^{2}}
\end{equation}
%
\begin{figure}
\epsfxsize= 8.0cm
\centerline{\epsffile[0 0 380 221]{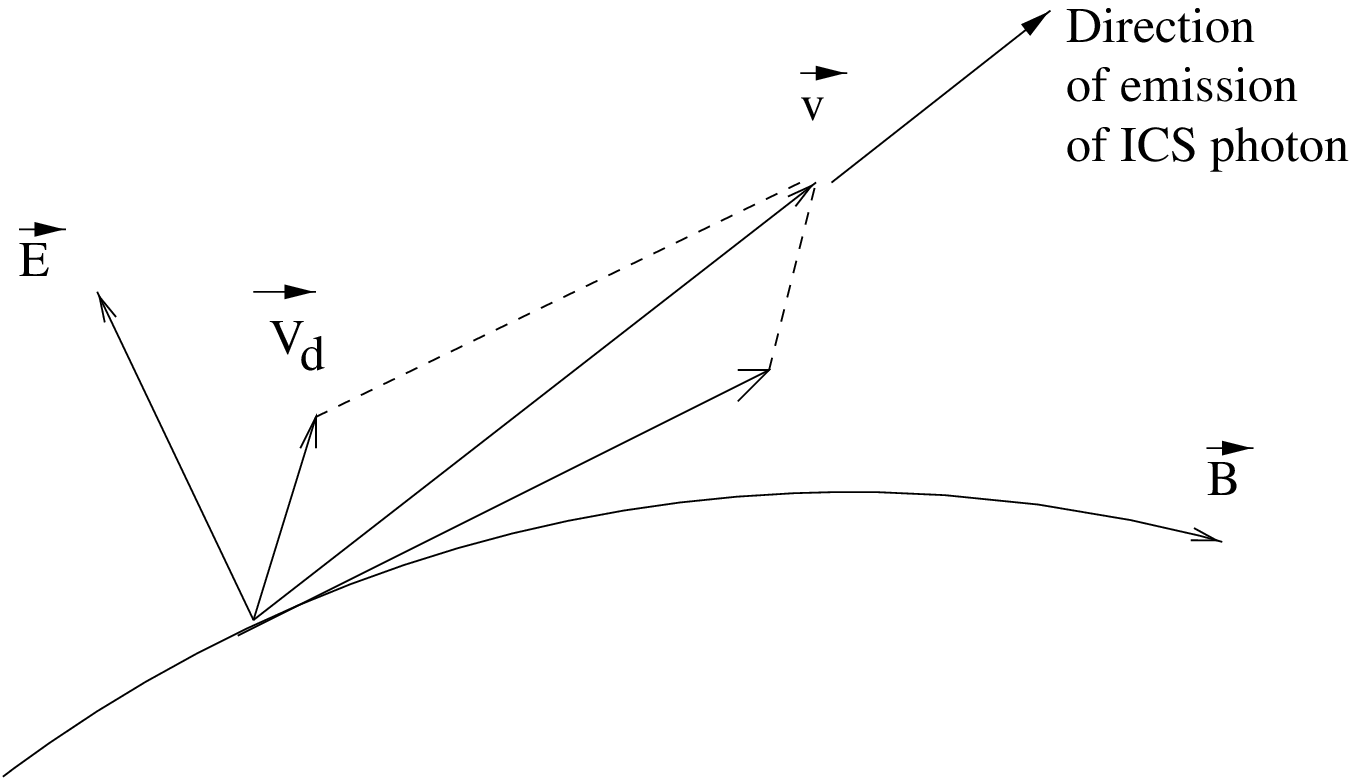}}
\caption{The geometry of the MHD plasma flow.}
\label{fig3}
\end{figure}

\begin{figure}
\epsfxsize= 8.0cm
\centerline{\epsffile[0 0 258 177]{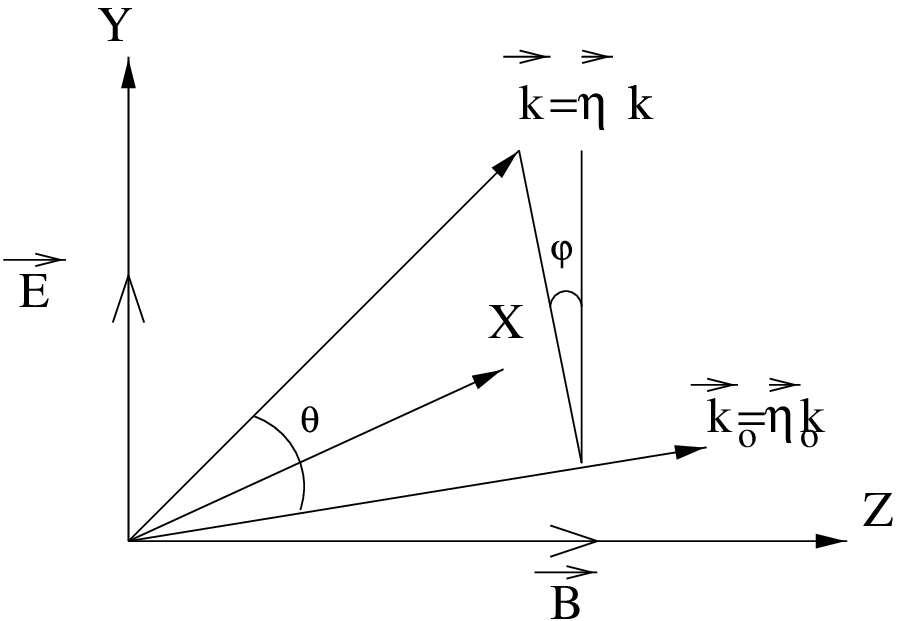}}
\caption{The geometry used in equation (\protect\ref{xi}).}
\label{fig4}
\end{figure}

 $V_{d}$  is comparable with c beyond  the light cylinder,
where the electric and
magnetic fields are of the same order. The schematic
relationship between the magnetic and electric  
fields and the velocity of plasma is shown
in Fig.  \ref{fig3}. In this situation an  
energetic photon is emitted along the velocity vector 
of the particle at large angle to the magnetic 
field.  However,  this happens in the region of crossed
electric and magnetic fields,  where the coefficient 
of absorption of the photon is modified by the
electric field. The probability  of conversion of the photon on unit length,
taking into account the electric field,  was defined by Daugherty \& Lerche: 
\shortcite{daugherty}
\begin{equation}
\xi = 0.23\alpha{mc\over\hbar}{B\chi\over B_{cr}}
{(1-E^{2}/B^{2})\over (1-E\eta_{\rm x}/ B)} \
\exp{\left(-{8\over 3}{B_{\rm cr}\over E_{\gamma} \chi B}\right)},
\label{xi}
\end{equation}
where
\begin{equation}
\chi= \sqrt{(\eta_{x}-{E\over B})^{2}+\eta_{y}^{2}(1-{E^{2}\over B^{2}})},
\label{chi}
\end{equation}
and $\alpha={1/137}$. $\eta_{\rm x}$ and $\eta_{\rm y}$ are the components of the unit vector
directed along the velocity of the photon as it is shown in Fig. \ref{fig4} and $E_{\gamma}$ is the
energy of photon expressed in units of $mc^{2}$.

The components of the velocity of the particle in the system of coordinates presented
in Fig. \ref{fig4} are 
as follows $v_{x}={cE\over B}$, $v_{y}=0$, $v_{z}=c\sqrt{1-{E^{2}\over B^{2}}}$.
The corresponding components  of the unit vector 
${\bf \eta}$ of the photon at the place of the emission
will be $\eta_{\rm x}={v_{\rm x}\over v}={Ec\over vB}$, $\eta_{y}=0$,
$\eta_{\rm z}=\sqrt{1-\eta_{\rm x}^{2}}$. 
Substitution of these components in equation (\ref{chi})
gives us
\begin{equation}
\chi={E\over B}{c\over \gamma^{2}(1+v/c)v}
\end{equation}
This factor  is extremely small, $\chi \sim 10^{-14}$, for the expected parameters of the wind .

Photons propagate along straight lines. Charged particles move in electromagnetic fields and
 their trajectories diverge from straight lines.
If photon would moved along the trajectory of the charged 
particles, they would never convert in pairs.
Therefore it is clear that the probability of conversion of a  photon into a pair basically
depends on the
radius of curvature $R_{c}$ of the trajectory of the charged particles,
but not the magnetic field. 

The angle $\vartheta$ between the 
direction of propagation of the photon and direction of propagation of the
emitting electron depends on the path length of the photon $l$;
\begin{equation}
\vartheta =  {l\over R_{c}}.
\end{equation}

The dependence of components of the vector $\eta$ on $\theta$ is as follows
\begin{equation}
\eta_{x}= \eta_{x0}\cos{\vartheta}+\eta_{z0}\sin{\vartheta}\sin{\varphi}
\end{equation}
\begin{equation}
\eta_{y}= \sin{\vartheta}\cos{\varphi}
\end{equation}
Neglecting by terms of the order $1/\gamma^{2}$ and $\sin^{n}{\vartheta}$ 
in powers higher than 2, 
we obtain the following estimate for $\chi$ at small $\theta$
\begin{equation}
\chi \sim {l\over R_{\rm c}} \sqrt{1-\left({E\over B}\right)^{2}}.
\end{equation}
As  we are interested in the maximal values of $\chi$, we can neglect  the
term $\sqrt{1-({E\over B})^{2}}$ in this expression. 
Assuming that after the light cylinder the total magnetic field
decreases  as $1/r$, we obtain for the function 
$q={8B_{\rm cr}\over 3E_{\gamma}B\chi}$ the following
estimate of the upper limit
\begin{equation}
q={8B_{\rm cr}R_{\rm c}\over 3E_{\gamma} B_{\rm lc} r_{\rm L}},
\end{equation}
and for the probability of conversion of the photon 
into a  pair we obtain finally the estimate
\begin{equation}
\xi \le 0.23 \alpha {mc\over\hbar} {B_{\rm lc} r_{\rm L}
\over B_{\rm cr}R_{\rm c}}
\exp{(-{8B_{\rm cr}R_{\rm c}\over 3E_{\gamma} B_{\rm lc}r_{\rm L}})}
\end{equation}
where $B_{\rm lc}$ is the magnetic field on the 
light cylinder and  $R_{\rm c}$ is
also taken on the light cylinder.

The solution of the problem of the
relativistic  wind flows in the model of an axisymmetrically rotating star allows
us  to estimate the radius of curvature $R_{\rm c}$ of
the trajectories of the particles
in the wind \cite{bog97} as
\begin{equation}
{1\over R_{\rm c}}= {\sigma_{\rm w}\over \gamma^{2} r} \ .
\end{equation}

This estimate shows  that the radius of curvature of the 
trajectories of particles is
much more than the curvature radius of the magnetic field line, 
which is of the order
$r$. Such a large curvature radius  gives $q \gg 1$ for almost any
parameters of the wind. This  is why the IC photons produced in the wind 
are not converted into pairs even at
energies $\sim 100$ TeV
and $\sigma_{\rm w}\sim 1$.

\end{document}